\documentclass{collective-intelligence}

\usepackage{graphicx}
\usepackage[table]{xcolor}

\usepackage{balance}

\begin{document}

\title{A COMPUTATIONAL ANALYSIS OF COLLECTIVE DISCOURSE}
%
%
%
%
%

\numberofauthors{2} 
%
\author{
%
%
\alignauthor
{\bf Vahed Qazvinian}\\
       \affaddr{Department of EECS}\\
       \affaddr{University of Michigan}\\
       \affaddr{Ann Arbor, MI, 48109}\\
       \email{vahed@umich.edu}
\alignauthor
{\bf Dragomir R. Radev}\\
       \affaddr{School of Information}\\
       \affaddr{Department of EECS}\\
       \affaddr{University of Michigan}\\
       \affaddr{Ann Arbor, MI, 48109}\\
       \email{radev@umich.edu}
}

\maketitle
\begin{abstract}
This paper is focused on the computational analysis of collective discourse, a collective behavior seen in non-expert content contributions in online social media. We collect and analyze a wide range of real-world collective discourse datasets from movie user reviews to microblogs and news headlines to scientific citations. We show that all these datasets exhibit diversity of perspective, a property seen in other collective systems and a criterion in wise crowds. Our experiments also confirm that the network of different perspective co-occurrences exhibits the small-world property with high clustering of different perspectives. Finally, we show that non-expert contributions in collective discourse can be used to answer simple questions that are otherwise hard to answer.
\end{abstract}

\section{Introduction}

Collective behavior refers to social processes that are not centrally coordinated and emerge spontaneously ~\cite{blumer1951}. This definition distinguishes collective behavior from group behavior in a number of ways: (a) collective systems involve limited social interactions, (b) membership is fluid, and (c) it generates weak and unconventional norms~\cite{smelser1963}. Collective behavior is normally characterized by a complex system~\cite{mill&page2007}. 
A complex system is a system composed of interconnected parts (agents, processes, etc.) that as a whole exhibit one or more properties called emergent behavior. The emergent behavior, which is not obvious from the properties of the individuals, is called to be nonlinear (not derivable from the summations of the activity of individual components).

Nonlinear behavior has been widely observed in nature in the past. 
\citeasnoun{Gordon-99} explains how harvester ants achieve task allocation without any central control and only by means of continual adjustment. 
Moreover she argues that the cooperative behavior in the ant colony merely results from local interactions between individual ants and not a central controller. 
For instance, in ant colonies individual members react to local stimuli (in the form of chemical scent) depending only on their local environment. 
In the absence of a centralized decision maker, ant colonies exhibit complex behavior to solve geometric problems like shortest paths to food or maximum distance from all colony entrances to dispose of dead bodies.

Self-organized behavior is not specific to ants. 
Schools of fish, flocks of birds, herd of ungulate mammals are other examples of complex systems among animal groups~\cite{fisher2009}. Similarly pedestrians on a crowded sidewalk exhibit self-organization that leads to forming lanes along which walkers move in the same directions \cite{boccara2010modeling}.
It is argued that all examples of complex systems exhibit common characteristics:
\begin{enumerate}
\item They are composed of a large number of inter-connected parts (i.e., agents)
\item The system is self-organized in that there is not central controller.
\item They exhibit emergent behavior: properties seen in the group but not observable
\end{enumerate}

In social sciences, a lot of work has been done on collective systems and their properties~\cite{page2009}. 
However, there is only little work that studies a collective system in which individual members collectively describe an event or an object. In our work, we focus on the computational analysis of collective discourse, a collective behavior seen in interactive content contribution in online social media~\cite{vahed&radev2011}.

In this paper, we show that collective discourse exhibits diversity of opinions, a property that is defined by~\cite{Surow:2004} as a necessary criterion for wise crowds.

\section{Background}
Previously, it has been argued that diversity is essential in intelligent collective decision-making. \citeasnoun{page2007} argues that the diversity of people and groups, which enable new perspectives, leads to better decision making. 
He finds that the diversity of perspectives in a collective system is associated with higher rates of innovation and can enhance the capacity for finding solutions to complex problems. 
Similarly,~\citeasnoun{hong2004} show that a random group of intelligent problem solvers can benefit from diversity and outperform a group of the best problem solvers.

Prior work has also studied the diversity of perspectives in content contribution and text summarization. In prior work on evaluating independent contributions in content generation, ~\citeasnoun{Voorhees98} studied IR systems and showed that relevance judgments vary significantly between humans but relative rankings are more stable across annotators. 
Similarly, ~\citeasnoun{teufel&Halteren2004} designed an experiment, which asked 40 Dutch students and 10 NLP researchers to summarize a BBC news report, resulting in 50 different summaries. They calculated the Kappa statistic ~\cite{Carletta96,krippendorff} and observed high inter-judge agreement, suggesting that the task of atomic semantic unit (factoid) extraction can be robustly performed in naturally occurring text.

The diversity of perspectives and the unprecedented growth of the factoid inventory have influenced other research areas in Natural Language Processing such as text summarization and paraphrase generation. Summarization evaluations are performed by assessing the information content with respect to salience and diversity in the summaries that are generated automatically~\cite{sparck-jones-99,vanHalteren2003,nenkova2004ecs}.

Leveraging the diverse range of perspectives has also played a critical role in developing new paraphrase generation systems by providing massive amounts of data that is easily collectable. For instance,~\citeasnoun{chen-dolan} performed a study and collected highly parallel data, used for training paraphrase generation systems from descriptions that participants wrote for video segments from YouTube. Such parallel corpora of document pairs that represent the same semantic information in different languages can be extracted from user contributions in Wikipedia and be used for learning translations of words and phrases \cite{yih&al:2011}.

\section{Collective Discourse}
With the growth of Web 2.0, millions of individuals involve in collective discourse. 
They participate in online discussions, share their opinions, and generate content about the same artifacts, objects, and news events in Web portals like amazon.com, epinions.com, imdb.com and so forth. 
This massive amount of text is mainly written on the Web by non-expert individuals with different perspectives, and yet exhibits accurate knowledge as a whole.

In social media, collective discourse is often a collective reaction to an event. 
A collective reaction to a well-defined subject emerges in response to an event (a movie release, a breaking story, a newly published paper) in the form of independent writings (movie reviews, news headlines, citation sentences) by many individuals. 
To analyze collective discourse, we perform our analysis on a wide range of real-world datasets.

\begin{table}
\centering
\begin{tabular}{lcc}\hline 
Dataset & \#clusters & average \#docs\\  \hline \hline
Movie reviews & 100 & 965\\
Microblogs & 15 & 110 \\
News headlines & 25 & 55\\
Citations & 25 & 52 \\ \hline
\end{tabular}
\caption{Number and size of collective discourse datasets studied in this paper.}\label{tbl:data}
\end{table}

\subsection{Corpus Construction}
An essential step and an important contribution in our work is gathering a comprehensive corpus of datasets on collective discourse. We focus on social media consisting of independent contributions of many individuals. Furthermore, we focus on topics corresponding to specific items and events as opposed to issues that are evolving and diffuse either in time or scope such as the economy or education.  
Table~\ref{tbl:data} lists the set of collective discourse corpora that we have analyzed as well as the number of datasets and average number of documents in each of them. In the following, we further explain each of these collective discourse corpora.

\subsubsection{Movie Reviews}
The first collective discourse that we are interested in analyzing is the set of reviews that non-expert users write about a movie. The set of online reviews about an object is a perfect case of collective human behavior. Upon its release, each movie, book, or product receives hundreds and thousands of online reviews from non-expert Web users. These reviews, while discussing the same object, focus on different aspects of the object. For instance, in movie reviews, some reviewers solely focus on a few famous actors, while some discuss other aspects like music or screenplay.

To study collective discourse in movie reviews, we collected all the user reviews for 100 randomly
selected movies from the top 250 movies list in the Internet Movie Database (IMDB)\footnote{http://www.imdb.com/chart/top}. For each of these 100 movies, we also obtained plot keywords provided on the IMDB website. Our collected corpora consist of more than 96,500 user reviews posted for movies from 19 different genres.

The following excerpts are extracted from user reviews for the movie Pulp Fiction, and show how non-expert reviewers focus on different aspects of the movie.

\begin{description}
\item {\it ``... starred by many well-known Actors, such as: John Travolta, Samuel L. Jackson, Uma Thurman, Bruce Willis and many. Directed by Quentin Tarantino, the eccentric Director ...''}
\item {\it ``... Pulp fiction was nominated for seven academy awards and won only one for screen writing ...''}
\item {\it ``... Shocking, intelligent, exciting, hilarious and oddly though-provoking. Best bit: Jackson's Bible quote ...''}
\end{description}

\subsubsection{Microblogs}
The second type of collective discourse that we study in our work is the set of tweets written about a news story. 
In addition to other advantages, using Twitter as a corpus of collective discourse does present unusual challenges. 
In Twitter, posts are limited to 140 characters and often contain information in an unusually compressed form.

First, we use the set of tweets collected by~\cite{qazvinian&al:2011c} about Sarah Palin's divorce rumor that was popular during the 2008 presidential election campaigns. 
This dataset contains tweets that are about this story and yet discuss it from different angles. 
For example, the following tweets are extracted from this dataset and reveal various facts about the story. 
One aspect is that a blogger has started the spread, and is threatened with libel suit. 
Another aspect is that the rumor has been debunked on Facebook.

\begin{description}
\item {\it ``Palins lawyer threatens divorce blogger with libel suit, gives her the option of receiving the summons at her resid... http://ow.ly/15JDO6.''}
\item {\it ``@jose3030 Palin divorce is supposedly debunked on Facebook, but I think they are just spinning it, until they can announce it.''}
\item {\it ``RT @mediaite: Sarah Palin uses Facebook to deny unsourced divorce rumors - http://bit.ly/14Xy6h CH.''}
\end{description}

As our second Microblog dataset, we collected the tweets that talk about the cancellation rumors of 14 TV shows in August of 2011. For instance, one of our collected datasets is about the rumor that Charlie Sheen might go back to the TV show Two and a Half Men.

\begin{description}
\item {\it ``Charlie Sheen Claims 'Discussions' About Returning to 'Two and a Half Men': In Boston for his national tour, C... http://bit.ly/hIbOWf.''}
\item {\it Charlie Sheen ``Two And A Half Men'' Return Not Happening: Report http://dlvr.it/LCTkd.''}
\end{description}

\subsubsection{News Headlines}
Another collective discourse is seen when a story breaks and various news agencies write stories about it. These stories all talk about the story, but view it from different perspectives.

We collected 25 news clusters from Google News2. Each cluster consists of a set of unique headlines about the same story, written by different sources. The following example shows 3 headlines in our datasets that are about hurricane Bill and its damage in Maine.

\begin{description}
\item {\it ``Hurricane Bill sweeps several people into ocean.''}
\item {\it ``7-year-old girl swept away by Bill wave dies after rescue.''}
\item {\it ``Maine ranger: wave viewers didn't heed warnings.''}
\end{description}

\subsubsection{Citation Sentences}
The final collective discourse example that we study is the set of citation sentences that different scholars write about a specific paper. A citation sentence to an article, $P$, is a sentence that appears in the literature and cites $P$. Each citation to P may or may not discuss one of $P$'s contributions.

For example, the following set of citations to Eisner's work~\cite{eisner1996three} illustrate the set of factoid about this paper and suggest that different authors who cite a particular paper may discuss different contributions (fatoids) of that paper. 

\begin{description}
\item {\it In the context of DPs, this edge based factorization method was proposed by (Eisner, 1996).} 
\item  {\it  Eisner (1996) gave a  generative model with a  cubic parsing algorithm based on an edge factorization of trees.}
\item  {\it Eisner (1996) proposed an $O(n^3)$ parsing algorithm for PDG.}
\item  {\it If the parse has to be projective, Eisner's bottom-up-span algorithm (Eisner, 1996) can be used for the search.} 
\end{description}

\subsubsection{Other Collective Discourse Datasets}
The study of collective discourse helps us understand new aspects of an object that are hard to identify with a single authoritative view. Collective discourse examples are not limited to the datasets that we have collected. 
For instance, studying a complete set of introductions about PageRank enables us to learn about its important aspects such as the algorithm, the damping factor, and the Power method, as well as aspects that are less known such as its use in 1940s~\cite{Franceschet:2010:PSS}.
Similar examples exist in different TV show synopsis, book descriptions, story narrations and many more.

\section{Diverse Perspectives}
In social sciences, a \emph{perspective} is defined as a map from reality to one's internal language, which is used to describe millions of objects, events, or situations~\cite{page2007}. 
Each word in the internal language refers to a concept (factoid) that can be expressed by means of a spoken language using various words or phrases (nuggets). 
More accurately, a factoid is an atomic semantic unit, which can be represented using different phrasal information units or nuggets \cite{vahed&radev2011,vanHalteren2003}.

For instance, the ``death of a 7-year-old girl'' and ``kid, 7, dies'' are the same factoids about the hurricane Bill story but represented differently (using different nuggets). Sweeping several people and warnings before the hurricane are some other factoids in the set of headlines about this story. These factoids show that different news reporters focus on different aspects of the hurricane story. Similarly, ``Sarah Palin using Facebook to debunk the rumor'' is a factoid in the microblog dataset, and ``a bible quote mentioned by Samuel Jackson'' is a factoid that appears in the movie reviews about Pulp Fiction.

\subsection{Annotations}

For each collective discourse dataset, we construct the set of factoids that represent various aspects of a story or a movie or different contributions of a paper.

For the microblogs dataset, we asked two annotators to go over all the tweets and identify a set of factoids that represent different aspects of each rumor. We then manually marked each tweet with the factoid that is relevant to the tweet. 
Each factoid is usually covered by a number of tweets, and each tweet covers one or more factoids. However, we did not observe any tweets that cover more than 2 factoids in our datasets. 
The small number of factoids covered by each tweet is most likely due to the length limit enforced by Twitter on each post.

\begin{table}
\centering
\begin{tabular}{lcl}\hline
Factoid & \#tweets & Perspective description\\ \hline \hline
FB & 414 & debunked on Facebook\\
FAMILY & 106 & family values\\
ALASKA & 87 & Alaska report's evidence\\
QUIT & 72 & resignation and divorce\\
AFFAIRS & 58 & affairs\\
GAY & 36 & gay marriage ban\\
CAMP & 36 & her camp denies the rumor\\
MONTANA & 33 & moving to Montana\\
LIBEL & 24 & libel suit against the rumor\\
BLOG & 19 & blogger who started the rumor\\ \hline
\end{tabular}
\caption{Different factoids extracted from the Palin dataset with the number of tweets that mention them, and short descriptions.}\label{tbl:fb}
\end{table}

Table~\ref{tbl:fb} lists the factoids extracted from the Sarah Palin.s divorce rumor dataset. This table shows that the 414 tweets discuss how ``Facebook is used to debunk the rumor,'' while the ``libel suit against the blogger who started the rumor'' is only mentioned in 24 tweets of the total 789 tweets.

To calculate the inter-judge agreement, we annotated 100 microblog instances on Sarah Palin twice, and calculated the statistic as 
\begin{equation*}
\kappa = \frac{Pr(a) - Pr(\epsilon)}{1 - Pr(\epsilon)}
\end{equation*}
where $Pr(a)$ is the relative observed agreement among the two annotators on the 10 factoids from Table~\ref{tbl:fb}, and $Pr(\epsilon)$ is the probability that annotators agree by chance if each annotator is randomly assigning categories. Based on this formulation, we reach a value of 0.913 in $\kappa$, and $93\%$ agreement between the two annotators.

We also annotated the set of citations and news headlines in the same fashion. 
Particularly, we asked two annotators to extract factoids for each of the 25 news and citation clusters, and then match individual documents (headline or citation sentence) with relevant factoids. 
Previously we have shown high agreement in human judgments for extracting factoids from these datasets ($\kappa \approx 0.8$)~\cite{vahed&radev2011}.

For the movie review clusters, we downloaded the list of cast names as well as the list of plot keywords provided for each movie by IMDB, as the set of factoids about the movie\footnote{We admit that the set of cast names and plot keywords provided by IMDB does not include all the factoids about the movie. However, since creating gold standard data from complete user reviews is fairly arduous, and we did not pursue manual annotations for movies.}.

Table~\ref{tbl:numf} lists the average number of factoids for each collective discourse corpus. For the Movie reviews, there is an average of 131 factoids per movie, and for citations, headlines and microblogs, our annotators identify an average of 5, 7, and 3 factoids respectively.

\begin{table}
\centering
\begin{tabular}{lcr}\hline
Dataset & & Number of factoids\\ \hline \hline
Movie reviews & & $131.31 \pm 52.67$\\
Microblogs & & $2.93 \pm 2.05$\\
News headlines & & $7.48 \pm 4.02$\\
Citations & & $5.48 \pm 1.96$\\ \hline
\end{tabular}
\caption{Average number of factoids in various collective discourse corpora.}\label{tbl:numf}
\end{table}

\subsection{Diversity}
\citeasnoun{Surow:2004} defines 4 criteria for a crowd to be wise: (1) people in the crowd should have diverse knowledge of facts (diversity); (2) people should act independently and their opinion should not be affected by that of others (independence); (3) people should have access to local knowledge (decentralization); and (4) a mechanism should exist to turn individual judgments into collective intelligence (aggregation).

Here, we present evidence that the individuals who engage in collective discourse have diverse perspectives and interpret things differently.

\begin{figure}[ht!]
 \begin{center}$
 \begin{array}{cc}
\includegraphics[width=42mm]{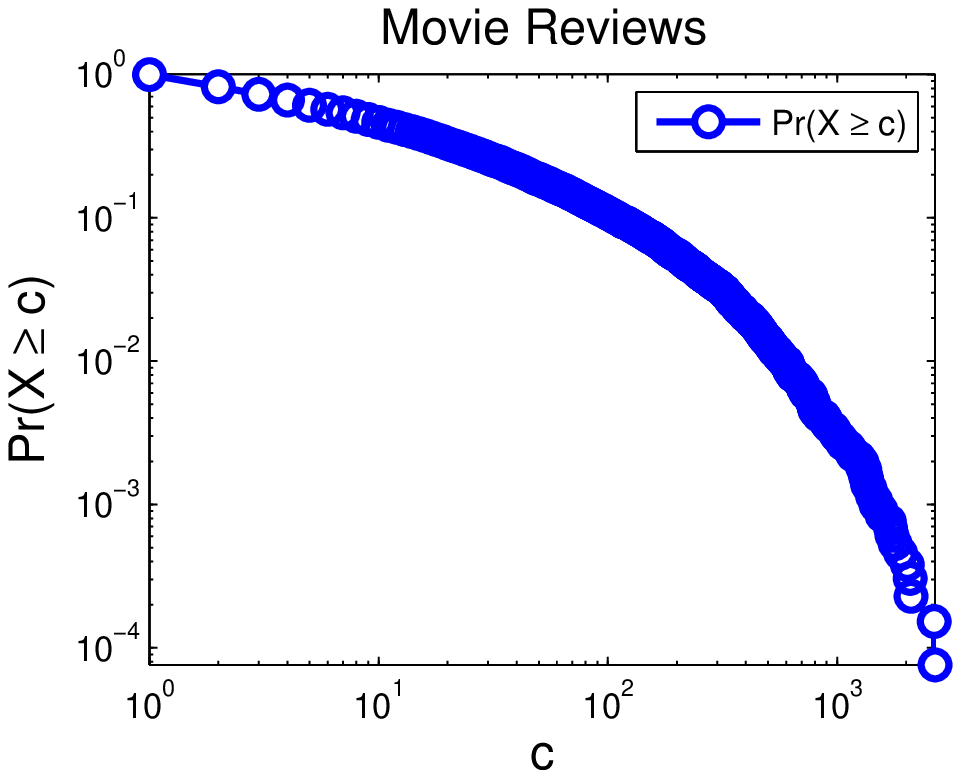} &
\includegraphics[width=42mm]{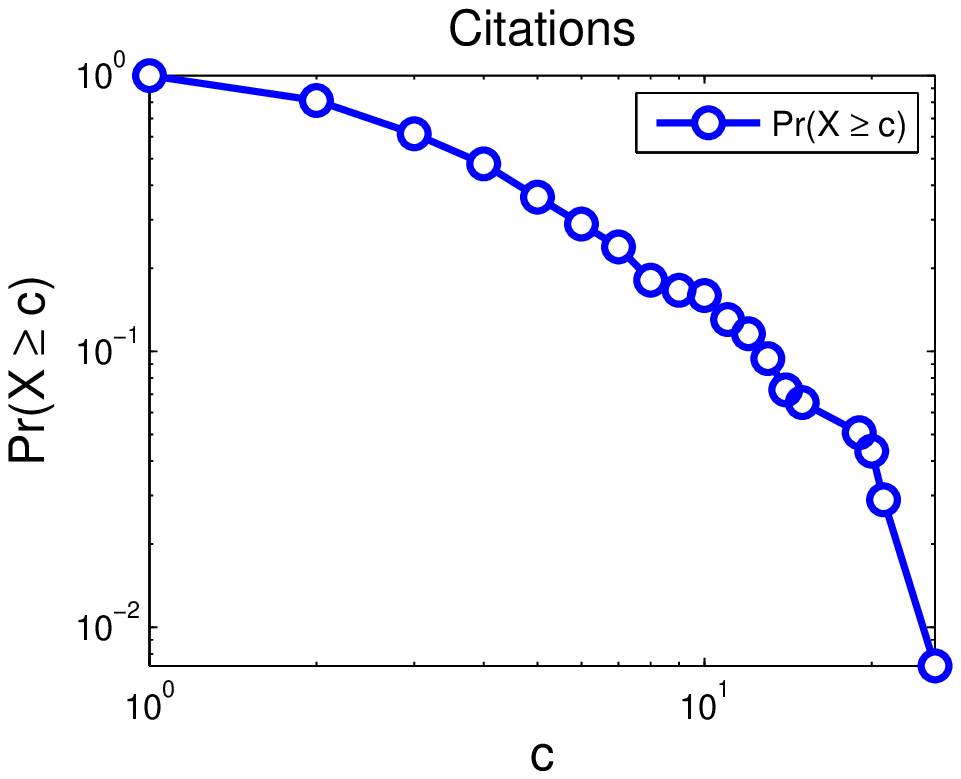} \\
\includegraphics[width=42mm]{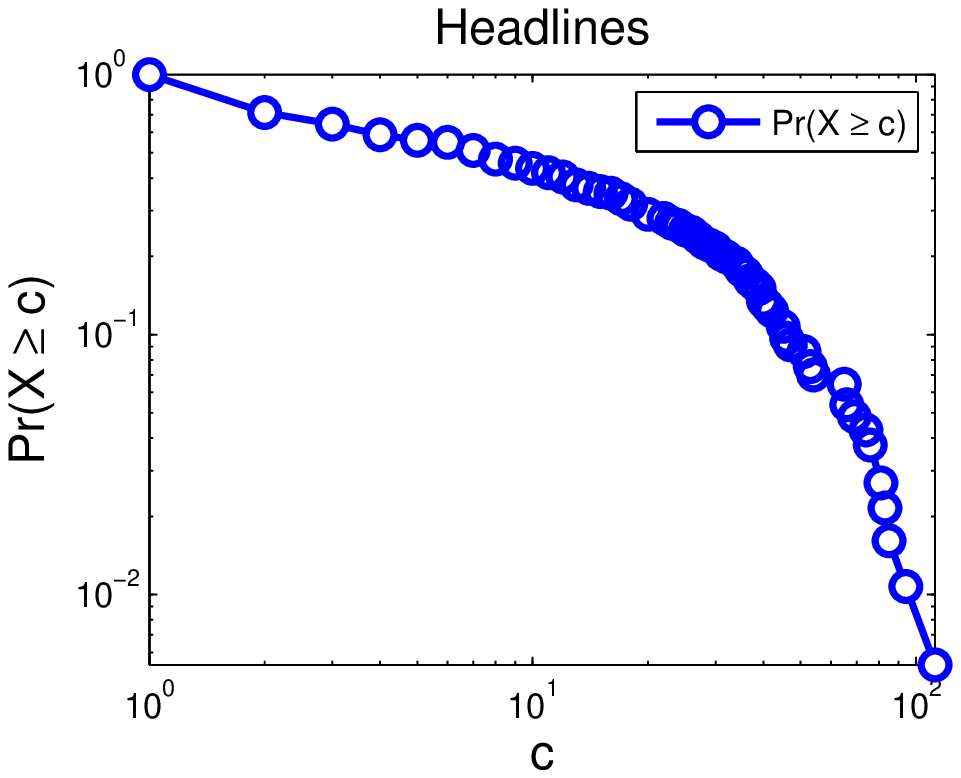} &
\includegraphics[width=42mm]{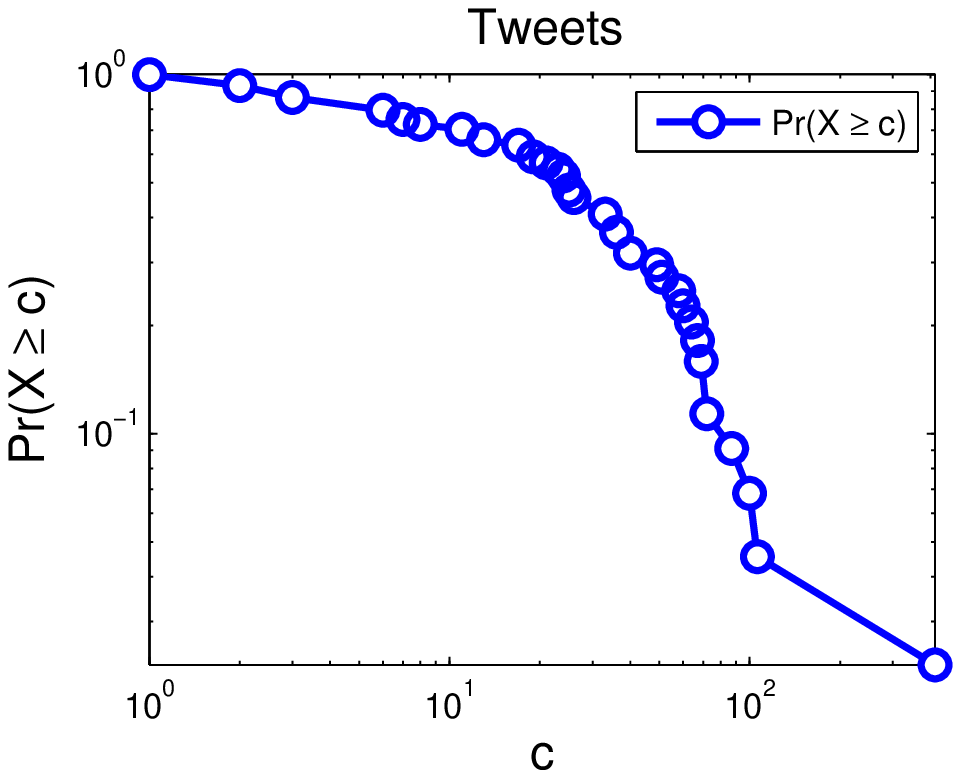} \\ 
 \end{array}$
\end{center}
 \caption{The cumulative probability distribution for the frequency of factoids (i.e., the probability that a factoid will be mentioned in $c$ different summaries) across in each corpus.}\label{fig:div}
\end{figure}

\subsubsection{Novelty and Redundancy}

To investigate the diversity of perspectives, we look at the frequency distribution of various factoids in different corpora by extracting the number of individuals that mention each factoid, $f$, in the annotated clusters. 
Figure~\ref{fig:div} shows the log-log scale cumulative probability distribution for these counts (i.e., the probability that a factoid will be mentioned by at least $c$ different people) in all of our collective discourse corpora. 
This figure suggests that factoid mention frequencies exhibit a highly skewed distribution with many factoids mentioned only once and a very few factoids mentioned by a large number of people. 
For instance, in the Pulp Fiction example, ``Bruce Willis'' and ``Quentin Tarantino'' are very popular factoids and most reviewers mention them, while ``Rene Beard'', ``Frank Whaley'' (two other actors), or ``Jackson's bible quote'' are among many factoids that are not as frequently mentioned.

\section{Small-world of Factoids}
Recent research has shown that a wide range of natural graphs such as the biological networks~\cite{ravasz2002hierarchical}, food webs~\cite{montoya2002small}, electronic circuits~\cite{FerrerICancho&etal2001}, brain neurons~\cite{bassett2006small}, and human languages~\cite{FerreriCancho&Sole01} exhibit the small-world property. 
This common characteristic  can be detected from two basic statistical properties: the clustering coefficient $\mathcal{C}$ , and  the average shortest path length $\ell$.

The clustering coefficient of a graph measures the  number of closed triangles in the graph. 
The  clustering coefficient describes how likely it is that two neighbors of a vertex are  connected~\cite{Newman03a}.
\citeasnoun{Watts&Strogatz1998} define the  clustering coefficient  as the average of the  local clustering values for each vertex.

\begin{equation*}
\mathcal{C} = \frac{\sum_{i=1}^n c_i}{n}
\end{equation*}

The local clustering coefficient,  $c_i$ for the $i$th vertex is  the number of triangles connected to vertex $i$  divided 
by the total possible number of triangles connected to vertex. \citeasnoun{Watts&Strogatz1998} show that small-world networks are highly clustered and obtain 
relatively short paths (i.e., $\ell$ is small). 
These networks are usually studied in contrast with random networks in which both $\ell$  and $\mathcal{C}$  obtain small values.

To understand  the relationship between various
aspects of a story  or subject and to  study the 
relationship between  different individuals' 
contributions, we analyze the network of factoids. 

For each dataset, we build a network in which nodes 
represent different factoids and there is an edge 
between two nodes if the corresponding factoids have 
been mentioned  together  in  at least 10 documents.  
Using these  networks, we would like to investigate 
whether there are many factoid pairs that co-occur in 
individual user contributions, and whether  there are 
communities  of factoids that co-occur more 
frequently than others. For each network, we use the 
same number of nodes and edges  and generate  a 
random network using the  Erd{\"o}s--R{\'e}nyi
model, which sets an edge between each pair of nodes with 
equal probability, independently of the other edges~\cite{ErdosRenyi}.

Table \ref{tbl:sw} lists the  average clustering coefficient  ($\mathcal{C}$) 
and the average shortest path length ($\ell$) in the 
networks built using factoid co-occurrences. This 
table confirms that the clustering  coefficient in the 
factoid networks is generally significantly greater 
than random networks  of the same size.  Moreover,
this table confirms that the average shortest paths in 
the random networks are small.

\begin{table}
\centering
\begin{tabular}{cccc}\hline
$\mathcal{C}$ & $\mathcal{C}_{random}$ & $\ell$ & $\ell_{random}$\\ \hline \hline
0.814 &  0.072 & 1.613 & 2.627\\ \hline
\end{tabular}
\caption{Average  clustering  coefficient ($\mathcal{C}$) and the average shortest path length ($\ell$)  in the  networks of  the collective discourse 
corpora  and the corresponding random networks.}\label{tbl:sw}
\end{table}

\citeasnoun{FerreriCancho&Sole01} and \citeasnoun{Motter&al.02} perform similar experiments and show that the  word co-occurrence and word synonymy networks
have small-world properties.  However, we believe  that this is the first work that shows the small-world  effect in human language at the factoid level
(network of concepts). This finding further justifies the conclusion made by \cite{Motter&al.02}, who emphasize that human memory is associative (i.e., information is retrieved by connecting similar concepts) in which the small-world property of the network maximizes the retrieval efficiency. More 
precisely, high clustering of the network causes similar pieces of information to be stored together, and low shortest paths make very different pieces of information to be separated only by a few links, guaranteeing a fast search.

\section{Wise Crowds}
Previous work has studied crowd wisdom in online content contributions.  Wikipedia for instance, has been named as an example of a successful collective 
effort. ~\citeasnoun{kittur&al2007} study user contributions in Wikipedia and suggest that  the main workload is  Wikipedia is driven by  ``common'' users and  that admin influence has dramatically decreased over years. 
Furthermore, ~\citeasnoun{Kittur:2008} show that  adding more editors to an article  results in higher article quality when appropriate coordination techniques are used.
In this section, we present some evidence of wisdom  in collective discourse that is  not achievable  from individuals  or from smaller groups.  In our 
experiments, we try to answer  a simple question about a movie just by using its set of reviews.

The question we try to answer is to find each movie's genre. 
As the gold standard, we  collected the genres for each of the 100 movies for which we had user reviews.  Each  movie is associated with  a few (3-4) genres out of a total of 19 genre names.

To extract the list of possible genres for a movie, we match all the genre names against the reviews and  rank them based on  their relative frequency.
More particularly, the score of each genre, $g$  for a movie with $N$  reviews ($D_1 ... D_N$) is calculated as 

\begin{equation*}
S_g =  \frac{\sum_{i = 1}^N {\bf 1}_{D_i \textrm{ mentions }g}}{N}
\end{equation*}

Table~\ref{tbl:avatar} lists the top 10 genres retrieved for the movie 
``Avatar'' form user reviews together with the score of 
each genre and the relevance according to the gold 
standard that we obtained from IMDB. This table 
shows an example in which all the 4 genre names for 
Avatar  are among the 7 most frequently genres 
mentioned by non-expert users.

\begin{table}
\centering
\begin{tabular}{llcc}\hline 
Rank & Genre & $S_g$ & relevance \\ \hline \hline
\rowcolor{pink} 1 &  {\bf action}  & 0.241 & 1\\
\rowcolor{pink} 2 &  {\bf sci-fi}  & 0.124 & 1\\
3 &  war     & 0.105 & 0\\
\rowcolor{pink} 4 &  {\bf fantasy} & 0.087 & 1\\
5 &  history & 0.086  & 0\\
6 &  animation & 0.062 & 0\\
\rowcolor{pink} 7 &  {\bf adventure} & 0.051 & 1\\
8 &  romance   & 0.039 & 0\\
9  & drama     & 0.025 & 0\\
10 & family & 0.023 &  0\\\hline
\end{tabular}
\caption{Top 10 genres extracted for the movie ``Avatar'' from user reviews.}\label{tbl:avatar}
\end{table}

To evaluate the ranked list of retrieved genre names, 
we use  Mean Average Precision and F-score. The 
Mean Average Precision (MAP) for a set of queries 
(movie names in our experiments) is calculated as the 
mean of the average precision scores for each query. 
The average precision for each query, $q$  is calculated 
as

\begin{equation}
AP_q = \frac{\sum_{k=1}^N Precision@k \times rel(k)}{\textrm{number of relevant genres}}
\end{equation}

where $rel(k)$ obtains a value of 1 if the $k$th retrieved genre is correct and 0 otherwise. We also calculate $F_{\beta = 3}$ when top 3 genres from the top of the ranked list are retrieved as relevant. Table~\ref{tbl:rslt} lists the results of this experiment.

To see how useful the set of reviews is for this 
particular task, we compare it  with ranking genre 
names randomly and repeating the experiment. As 
Table 6 shows, using simple mention frequency 
measures provides significant improvements over 
guessing the genre randomly.

The numbers in Table~\ref{tbl:rslt} are calculated using all the 
user reviews collected for each movie (ranging from 
a few hundreds to a few thousands per movie). Here, 
we would like to investigate if having more reviews 
will give us a more accurate estimate of the genres 
associated with each movie. 

\begin{table}
\centering
\begin{tabular}{lcccc}\hline 
Method & MAP & $95\%$ C.I. & $F_{\beta = 3}$ & $95\%$ C.I.\\ \hline \hline
Reviews &  0.698  &  [0.657 , 0.740] & 0.550  & [0.499 , 0.600]\\
Random  &  0.260  &  [0.229 , 0.290] & 0.140  & [0.101 , 0.179]\\ \hline
\end{tabular}
\caption{Mean Average Precision and F-score for genre extraction from a set of reviews  (C.I.: Confidence Interval).}\label{tbl:rslt}
\end{table}

\begin{figure}[ht!]
\includegraphics[width=9cm]{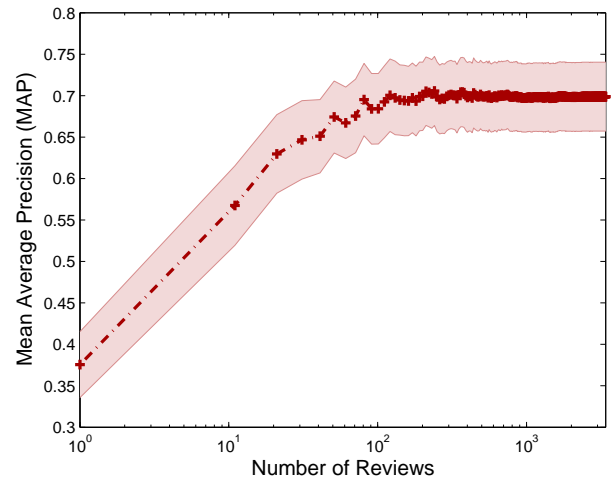}
 \caption{Mean Average Precision (MAP) versus the number of  reviews 
used to extract each movie genre. (The shaded area shows 95\% confidence interval for each MAP result)}\label{fig:curve}
\end{figure}

Figure~\ref{fig:curve}  plots the $95\%$ confidence interval of MAP
versus the number of randomly selected user reviews 
used to rank the genres for each movie. This figure, 
which is plotted on a semi-log scale, shows that the 
quality of  ranking  grows  rapidly  by  the  100th
randomly selected review and  exhibits asymptotic 
behavior when more reviews are visited.

\section{conclusion and future work}
We  studied collective discourse  and  investigated
diverse perspectives  when a number of non-expert 
Web users engage in collective behavior and generate 
content on the Web. We show that the set of people 
who discuss  the same story  or subject  have  diverse 
perspectives, introducing new aspects that have not 
been previously discussed by others.

We  analyzed a wide range of collective discourse 
examples, from movie reviews and  news stories to 
scientific citations and microblogs. To the best of our 
knowledge this is the first work  that studies the 
diversity in  perspectives, and the small world-effect 
in  factoid co-occurrences.  We also perform an 
experiment that provides some evidence of collective 
intelligence in the collectively written set of reviews 
by non-expert users.

The ultimate goal of this work is to develop models 
of collective discourse. The models would be 
informed by empirical analysis of varied and large-scale datasets and would address various aspects of 
collective discourse: motivation behind continuous 
contributions, heterogeneity and diversity  in 
perspectives, and collective intelligence from 
collaboration. By formulating simple stochastic 
models of individual and group behavior, we may be 
able to predict phenomena on the macro level of 
discourse.  We will be trying to  address these 
questions by developing state of the art technologies 
in computational linguistics, network science and 
social theories of mass communications.

\section{Acknowledgments}
This work is supported by the National Science
Foundation grants ``SoCS: Assessing Information Credibility Without Authoritative Sources'' as IIS-0968489, and ``iOPENER: A Flexible Framework to Support Rapid Learning in Unfamiliar Research Domains'' as IIS-0705832. Any opinions, findings, and conclusions or recommendations expressed in this paper are those of the authors and do not necessarily reflect the views of the National Science Foundation.

\balance
\bibliographystyle{agsm}
\bibliography{tref} 

\end{document}